\documentclass[twocolumn,superscriptaddress,floatfix,preprintnumbers, nofootinbib,hyperref]{revtex4-2} 
\pdfoutput=1
\usepackage[colorlinks=true,breaklinks=true]{hyperref}
\usepackage[normalem]{ulem}
\usepackage[utf8]{inputenc}
\hypersetup{allcolors=[rgb]{0.0 0.0 0.6},linkcolor=[rgb]{0.75 0.05 0.05}}
\usepackage{amsmath,amssymb}
\usepackage{epsfig}  
\usepackage{graphicx}   
\usepackage{slashed}       
\usepackage{tikz}
\usepackage{url}
\usepackage{color}
\usepackage{multirow}
\usepackage{comment}
\usepackage{amssymb}
\usepackage{orcidlink}

\DeclareMathOperator{\cm}{cm}
\DeclareMathOperator{\GeV}{GeV}

\DeclareMathOperator{\MeV}{MeV}

\DeclareMathOperator{\s}{s}

\DeclareMathOperator{\erg}{erg}

\DeclareMathOperator{\km}{km}

\DeclareMathOperator{\kpc}{kpc}

\DeclareMathOperator{\Mpc}{Mpc}

\newcommand{\bp}{{\bf p}}

\allowdisplaybreaks

\bibpunct{[}{]}{,}{n}{}{,}
\definecolor{ForestGreen}{RGB}{34,139,34}

\begin{document}

\title{Proto-neutron stars as cosmic factories for massive
axion-like-particles}

\author{Alessandro Lella \orcidlink{0000-0002-3266-3154}} 
\email{alessandro.lella@ba.infn.it}
\affiliation{Dipartimento Interateneo di Fisica  ``Michelangelo Merlin,'' Via Amendola 173, 70126 Bari, Italy}
\affiliation{Istituto Nazionale di Fisica Nucleare - Sezione di Bari, Via Orabona 4, 70126 Bari, Italy}%

\author{Pierluca Carenza \orcidlink{0000-0002-8410-0345}}\email{pierluca.carenza@fysik.su.se}
\affiliation{The Oskar Klein Centre, Department of Physics, Stockholm University, Stockholm 106 91, Sweden
}

\author{Giuseppe~Lucente~\orcidlink{0000-0003-1530-4851}}
\email{giuseppe.lucente@ba.infn.it}
\affiliation{Dipartimento Interateneo di Fisica  ``Michelangelo Merlin,'' Via Amendola 173, 70126 Bari, Italy}
\affiliation{Istituto Nazionale di Fisica Nucleare - Sezione di Bari, Via Orabona 4, 70126 Bari, Italy}%

\author{Maurizio Giannotti \orcidlink{0000-0001-9823-6262}}
\email{mgiannotti@barry.edu}
\affiliation{Department of Chemistry and Physics, Barry University, 11300 NE 2nd Ave., Miami Shores, FL 33161, USA}%

\author{Alessandro Mirizzi \orcidlink{0000-0002-5382-3786}} 
\email{alessandro.mirizzi@ba.infn.it}
\affiliation{Dipartimento Interateneo di Fisica  ``Michelangelo Merlin,'' Via Amendola 173, 70126 Bari, Italy}
\affiliation{Istituto Nazionale di Fisica Nucleare - Sezione di Bari, Via Orabona 4, 70126 Bari, Italy}%

\date{\today}
\smallskip

\begin{abstract}
The parameter space of massive axion-like-particles (ALPs) with $m_a \sim {\mathcal O} (100)$~MeV and coupled with nucleons is largely unexplored. Here, we present new constraints in this parameter region.  In doing so, we characterize the supernova emissivity of heavy ALPs from a proto-neutron star, including for the first time mass effects in both nucleon-nucleon Bremsstrahlung and pionic Compton processes. In addition, we highlight novel possibilities to probe the couplings with photons and leptons from supernova ALP decays.
\end{abstract}

\maketitle

\section{Introduction}
\label{sec:1}

Core-collapse supernovae (SNe) are powerful cosmic laboratories to probe the emission of light particles with masses $\lesssim {\mathcal O}$(100) MeV (see, e.g., Refs.~\cite{Raffelt:1990yz,Raffelt:1987yt} for reviews). Particularly relevant, in this respect, was the observation of a neutrino burst from the SN 1987A explosion~\cite{Kamiokande-II:1987idp,Hirata:1988ad,Bionta:1987qt}. Despite the sparseness of the recorded neutrino events in the existing experiments, this observation was a milestone event in astroparticle physics, allowing the setting of stringent bounds on standard and non-standard neutrino properties~\cite{Kolb:1987qy,Raffelt:1990yu}. Furthermore, the data were used to constrain other novel particles emission such as axions~\cite{Turner:1987by,Brinkmann:1988vi,Chang:2018rso,Lucente:2022vuo}, muonic bosons~\cite{Caputo:2021rux}, dark photons~\cite{DeRocco:2019njg}, gravitons~\cite{Hannestad:2001jv}, unparticles~\cite{Hannestad:2007ys}, scalars mixed with Higgs boson~\cite{Balaji:2022noj} and particles escaping in extra dimension~\cite{Friedland:2007yj}, since their emission might have shortened the observed duration of the neutrino burst. 

Arguably, the most studied cases in this context are those of QCD axions or generic axion-like-particles (ALPs) coupled with nucleons. In presence of a dominant nucleon coupling, the ALP emissivity in the SN nuclear medium is dominated by nucleon-nucleon ($NN$) Bremsstrahlung~\cite{Carena:1988kr,Brinkmann:1988vi,Raffelt:1993ix,Raffelt:1996wa,Carenza:2019pxu} $NN\to NNa$, and by pionic Compton-like scatterings $\pi^{-} p\to n a$, which convert a pion into an ALP~\cite{Turner:1991ax,Raffelt:1993ix,Keil:1996ju}. The characterization of these processes is still a field of intense investigations with significant developments. In particular, the naive approach for the $NN$ process was based on the One-Pion-Exchange (OPE) scheme, where the interaction was assumed to be mediated by a pion~\cite{Turner:1987by,Carena:1988kr,Brinkmann:1988vi}. However, it has been recognized that different corrections, namely a non-vanishing mass for the exchanged pion~\cite{Stoica:2009zh}, the contribution from the two-pions exchange~\cite{Ericson:1988wr}, effective in-medium nucleon masses and multiple nucleon scatterings~\cite{Raffelt:1991pw,Janka:1995ir} would significantly reduce the ALP emissivity (see  Ref.~\cite{Carenza:2019pxu} for a state-of-the-art calculation). 
On the other hand, after a few seminal papers discussed it in the late 80s and early 90s, the Compton pionic process ($\pi N$) was largely neglected until Ref.~\cite{Carenza:2020cis} pointed out that it may dominate over the $NN$ process for typical conditions in the SN core.

Traditionally, the ALP emissivity from a proto-neutron star (PNS) has been calculated neglecting the ALP mass. This assumption is valid as long as \mbox{$m_a \ll 3 T\approx 90$~MeV}. At larger masses, the rate is suppressed by the Boltzmann factor $e^{-m_a/T}$ and by the ALP velocity \mbox{$\beta_a\sim 3T/m_a$}. The ALP parameter space for ALPs with mass \mbox{$m_a \sim {\mathcal O} (100)$~MeV} and coupled with nucleon is very poorly constrained. Therefore, PNS would represent a unique source to constrain massive ALPs. A first attempt to include the effects of a finite ALP mass in the SN emissivity goes back to Ref.~\cite{Giannotti:2005tn}, which calculated the $NN$ process within the OPE scheme.
The goal of our work is to extend these results to include all the relevant corrections beyond OPE, and to compute, for the first time, the effects of the axion mass in the $\pi N$ process. Furthermore, we will present several phenomenological implications of our results, new constraints, and future prospects. 
 
The plan of our work is as follows. In Section~\ref{sec:2}, we characterize the ALP emissivity in a SN core via $NN$ and $\pi N$ process including the effect of the ALP mass. In Section~\ref{sec:3}, we investigate the phenomenological consequences of the ALP mass on the SN cooling bound and on the ALPs gravitationally trapped near the SN core. 
In the subsequent Sections, we extend our phenomenological analysis to ALPs coupled also to photons or to leptons and therefore unstable. In particular, in Section~\ref{sec:4}, we consider radiative decays that, depending on the decay length, would give additional bounds from energy-deposition in the SN or from the production of a gamma-ray flux. In Section~\ref{sec:5}, we develop analogous considerations referring to ALP decays into electrons or muons. Finally, in Section~\ref{sec:conclusions}, we summarize our results and we conclude.

\section{Mass effects on axion emissivity in nuclear matter}
\label{sec:2}

The ALP nucleon interaction is described by the following Lagrangian terms~\cite{DiLuzio:2020wdo,Chang:1993gm}
\begin{equation}
    \begin{split}
        \mathcal{L}_{\rm{int}}&=g_a\frac{\partial_\mu a}{2m_N}\Bigg[C_{ap}\Bar{p}\gamma^\mu\gamma_5p+C_{an}\Bar{n}\gamma^\mu\gamma_5n+\\
        &+\frac{C_{a\pi N}}{f_\pi}(i\pi^+\Bar{p}\gamma^\mu n-i\pi^-\Bar{n}\gamma^\mu p)+\\
        &+C_{aN\Delta}\left(\Bar{p}\,\Delta^+_\mu+\overline{\Delta^+_\mu}\,p+\Bar{n}\,\Delta^0_\mu+\overline{\Delta^0_\mu}\,n\right)\Bigg]\,,
    \end{split}
\label{eq:NuclearInteractions}
\end{equation}
where $g_a$ is a dimensionless constant characterizing the ALP coupling with nuclei,\footnote{In the literature, especially that concerning QCD axions, it is common to find this coupling expressed in terms of the axion decay constant, $f_a$. The relation between $g_a$ and $f_a$ is $g_a=m_N/f_a$.} $f_{\pi}=92.4~\MeV$ is the pion decay constant, $C_{a\pi N}=(C_{ap}-C_{an})/\sqrt{2}g_{A}$~\cite{Choi:2021ign} and ${C_{aN\Delta}=-\sqrt{3}/2\,(C_{ap}-C_{an})}$, with $g_{A}\simeq1.28$~\cite{Workman:2022ynf} the axial coupling. 
The second term in Eq.~\eqref{eq:NuclearInteractions} gives rise to a four-particle interaction vertex, i.e. a contact interaction, whose contribution to the ALP emissivity was originally discussed in Ref.~\cite{Carena:1988kr} and recently re-discussed in Ref.~\cite{Choi:2021ign}. 
Finally, the last term describes the ALP couplings to the $\Delta$-resonance which give rise to an enhancement of the ALP emissivity, as recently discussed in Ref.~\cite{Ho:2022oaw}. 
It is convenient to define the ALP couplings with protons and neutrons as $g_{aN}=g_a\, C_{aN}$, for $N=p,n$, where $C_{aN}$ are model-dependent coupling constant. To compare with the literature, in this work we will often refer to the benchmark values $C_{ap}=-0.47$ and $C_{an}=0$, inspired by the Kim-Shifman-Vainshtein-Zakharov (KSVZ) axion model~\cite{GrillidiCortona:2015jxo}. 

The interaction Lagrangian in Eq.~\eqref{eq:NuclearInteractions} allows one to characterize the ALP emissivity in a nuclear medium, where the dominant processes are $NN$ Bremsstrahlung, $NN\to NNa$~\cite{Carena:1988kr,Brinkmann:1988vi,Raffelt:1993ix,Raffelt:1996wa,Carenza:2019pxu}, and the Compton-like scattering, $\pi^{-} p\to n a$, which converts a pion into an ALP~\cite{Turner:1987by,Burrows:1988ah,Burrows:1990pk,Raffelt:1987yt,Raffelt:1990yz} (see also Refs.~\cite{Carenza:2019pxu,Carenza:2020cis} for recent developments).
As discussed in the previous section, traditionally the calculation of these two processes has been performed assuming that the ALP mass $m_{a}$ is negligible compared to the environment temperature $T$, i.e, $m_{a}\ll T$. In this case, one expects a negligible Boltzmann suppression, $e^{-m_a/T}$, of the ALP emissivity. A first attempt to relax the massless ALP hypothesis in the $NN$ process, within the OPE framework, was presented in Ref.~\cite{Giannotti:2005tn}. In our work, we generalize these results and extend them also to the case of pionic processes. At first we notice that in the ALP massive case, for an ALP with mass $m_a<m_N$, the spin-summed matrix element of the $NN$ Bremsstrahlung can be written as~\cite{Giannotti:2005tn}
\begin{equation}
    S\displaystyle\sum_{\rm{spins}} |\mathcal{M}|^2=\left(
    \frac{{|{\bf p}_a|}}{\omega_a}\right)^{2}|\mathcal{M}^{(0)}|^2\,,
    \label{eq:MatElementBrem}
\end{equation}
where $|\mathcal{M}^{(0)}|^2$ is the spin-summed matrix element in the massless case~\cite{Brinkmann:1988vi,Carenza:2019pxu}, $\omega_{a}$ and $|\bp_{a}|$ are the ALP energy and momentum, respectively. Note that the only difference with the massless ALP case is the ALP velocity factor $\beta_a={|{\bf p}_a|}/{\omega_a}$. Then we can extend the ALP spectrum per unit volume calculated in Eq.~(2.15) of Ref.~\cite{Carenza:2019pxu} for massless ALPs to
\begin{eqnarray}
    \left(\frac{d^2n_a}{d\omega_a\,dt}\right)_{NN}&=&\frac{g_a^2}{16\pi^2} \frac{n_B}{m_N^2}(\omega_a^2-m_a^2)^{\frac{3}{2}}\exp\left(-\frac{\omega_a}{T}\right) \nonumber\\
    &\times & S_\sigma\left(\frac{\omega_a}{T}\right)\,\Theta(\omega_a-m_a)\,,
\label{eq:QaNN}
\end{eqnarray}
where $n_{B}$ is the baryon density in the SN core, and the Heaviside-$\Theta$ function guarantees that the minimum ALP energy is given by its mass. In the previous expression, $S_{\sigma}$ is the nucleon structure function, including the nuclear part of the matrix element~\cite{Raffelt:1996wa}
\begin{equation}
        S_{\sigma}(\omega_a)=\frac{\Gamma_\sigma}{\omega^2+\Gamma^2}s(\omega_a/T)\,,
\label{eq:StructureFunc}
\end{equation} 
where $s(x)$ and the spin-fluctuation rate $\Gamma_\sigma$ are given in Ref.~\cite{Carenza:2019pxu}. The Lorentzian form of Eq.~\eqref{eq:StructureFunc} takes into account multiple scattering effects~\cite{Raffelt:1996wa} and the line-width $\Gamma$ is chosen in such a way to have a properly normalized structure function (see Ref.~\cite{Sawyer:1989nu}). In particular, this structure function contains all the corrections beyond the OPE approximation introduced in Ref.~\cite{Carenza:2019pxu} to take into account the possibility of two pions exchange effects~\cite{Ericson:1988wr}, nucleons multiple scatterings~\cite{Raffelt:1996wa} and the effective nucleon mass inside the SN core due to a many body potential~\cite{Iwamoto:1984ir,Martinez-Pinedo:2012eaj,Hempel:2014ssa}. We highlight that in Eq.~\eqref{eq:StructureFunc} we include also a contribution related to the contact interaction term. This term has a negligible effect on the Bremsstrahlung emissivity in the whole range of masses studied, in agreement with the conclusions of Ref.~\cite{Choi:2021ign} obtained in the massless ALP case. As previously argued in Ref.~\cite{Carena:1988kr}, this is an expected result since the term due to this additional diagram gives a contribution of higher order in $1/m_N$ that is suppressed.

The other contribution to the ALP emissivity comes from the pion conversion that, as recently shown in Refs.~\cite{Fischer:2021jfm,Carenza:2020cis} is competitive with the $NN$ Bremsstrahlung for typical SN conditions. We extend the results of Eq.~(11) in Ref.~\cite{Carenza:2020cis} to the case of massive ALPs, calculating the spectrum per unit volume of ALPs as
\begin{equation}
    \begin{split}
        &\left(\frac{d^2n_a}{d\omega_a\,dt}\right)_{N\pi}=\frac{g_{a}^2 T^{3.5}}{2^{1.5}\pi^5m_N^{0.5}}\left(\frac{g_A}{2f_\pi}\right)^2\left(\omega_a^2-m_a^2\right)^\frac{1}{2}\\
        &\times\,\mathcal{C}_a^{p\pi^-}\frac{\Theta(\omega_{a}-m_{\pi})\,\Theta(\omega_{a}-m_{a})}{\exp{\left(x_a-y_\pi-\hat{\mu}_\pi\right)}-1}\,(\omega_a^2-m_\pi^2)^\frac{1}{2}\\
        &\times\int_0^\infty dy\, y^2\frac{1}{\exp{\left(y^2-\hat{\mu}_p\right)}+1}\frac{1}{\exp{\left(-y^2+\hat{\mu}_n\right)}+1}\,,
    \end{split}
    \label{eq:QaPionDef}
\end{equation}
where $m_{\pi}$ is the pion mass, $y_{\pi}=m_{\pi}/T$, $x_{a}=\omega_{a}/T$, $\hat{\mu}=(\mu-m)/T$, with $\mu$ the pion and nucleon chemical potentials, $\Theta$ is the Heaviside theta function, required to fix the minimal threshold energy at the ALP mass, and $\mathcal{C}_a^{p\pi^-}$ can be written as
\begin{equation}
    \mathcal{C}_a^{p\pi^-}=\frac{m_N^2}{g_A^2}\,\beta_a^2 \,\mathcal{G}_a(|\mathbf{p}_\pi|)\,,
\end{equation}

where $\mathcal{G}_a(|\mathbf{p}_\pi|)$ is taken from Eq.~(39) of Ref.~\cite{Ho:2022oaw}, neglecting terms of higher order in $1/m_N$. 
Notice that in the computation of the squared matrix element for massive ALPs we found a factor of difference $\beta_a^2$ w.r.t. to the massless case~\cite{Choi:2021ign,Ho:2022oaw}. 
In addition, we highlight that $\Delta$-mediated and contact interactions give rise to terms of the same order as the contribution coming from nucleon-mediated diagrams. 
Therefore, in agreement with what found in Ref.~\cite{Choi:2021ign,Ho:2022oaw}, we observe that each of these two couplings enhances the ALP emissivity by at least a factor of $\sim2$ in the mass range of interest $m_a\lesssim 500~\MeV$.\newline
These results are obtained employing a pion chemical potential and a pion abundance deduced from the procedure in Ref.~\cite{Fischer:2021jfm}, including the pion-nucleon interaction as described in Ref.~\cite{Fore:2019wib}.\newline

\section{PHENOMENOLOGICAL CONSEQUENCES}
\label{sec:3}

\subsection{ALP Energy spectrum}
\label{subsec:ALPspectrum}

The ALP emission is affected by the gravitational field of the PNS, in particular time dilation and the red-shifting of the energy~\cite{Caputo:2022mah,Calore:2021hhn}. In particular, the observed energy at infinity $\omega_{a}^{*}$ is shifted with respect to the local energy $\omega_{a}$ as 
\begin{equation}
    \omega_{a}^*=\alpha(r)\,\omega_{a} \,,
    \label{eq:redsh}
\end{equation}
where $\alpha(r)\leq 1$ is the so-called ``lapse factor'', which encodes effects due to the PNS gravitational potential $\Phi(r)$, evaluated locally at the PNS interior. The values of $\alpha$ at each radius are obtained from the SN simulation, e.g. Ref.~\cite{Kuroda:2021eiv}. Similarly, for a local observer, time dilation must be taken into account as follows
\begin{equation}
dt^*= \alpha^{-1}(r) \,dt(r)\,,   
\end{equation}
where the $dt(r)$ refers to the local observer interval of time at distance $r$, while $dt^*$ refers to the simulation time corresponding to that of a distant observer. Since \mbox{$d\omega_{a}^{*} dt^{*}= d\omega_{a}  dt$}, the integrated ALP spectrum over the local time $\alpha \,dt^*$ is given by
\begin{equation}
    \begin{split}
    \frac{dN_{\rm a}}{d \omega_a}
     &= \int d^{3}r dt^{*} \alpha(r) \frac{d{n}^{*}_{\rm a}}{d {\omega^{*}_a} dt^{*}}=\int d^{3}r dt \frac{dn_{\rm a}}{d\omega_a dt} \,\ .
    \end{split}
\end{equation}

\begin{figure} [t!]
\centering
	\includegraphics[width=1\columnwidth]{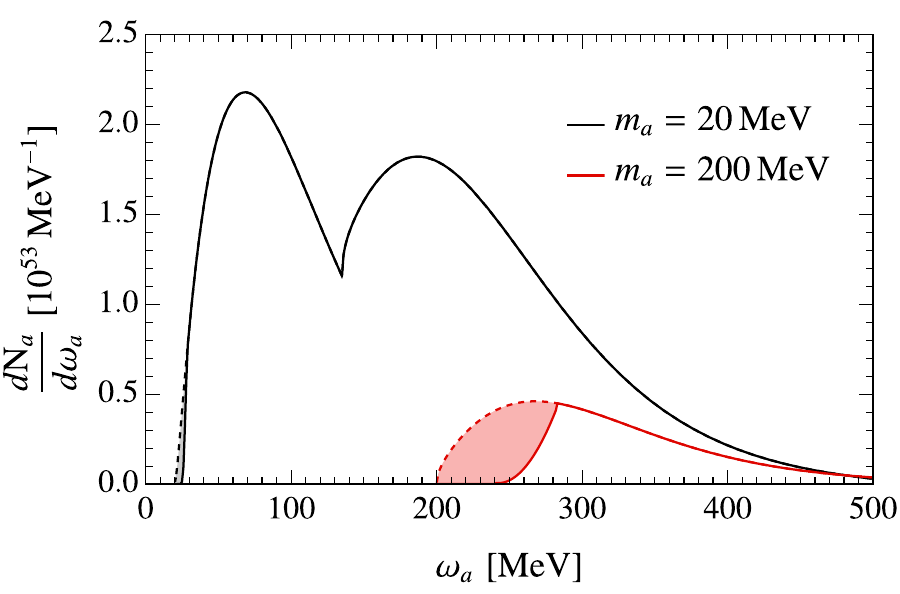}
	\caption{ALP energy spectra for
	$g_{ap}=10^{-10}$ and 
	two representative  masses,
	$m_a=20$~MeV (black curve) and $m_a=200$~MeV (red curve), integrated over $10$~s after the core bounce. Continuous curves are obtained including gravitational effects, while in dashed curves  the lapse function $\alpha=1$.
	The spectrum of gravitationally trapped ALPs is in the shaded bands.}
	\label{fig:Spectra}
\end{figure}

ALPs with sufficient kinetic energy can escape the stellar gravitational field. Relativistically, this condition can be expressed as $\omega_a > m_a/\alpha$ (see. e.g., Ref.~\cite{Caputo:2022mah}). In the weak gravitational field limit and for non-relativistic ALPs, this condition reduces to the well known result $\beta_a^2> 2M/r$ (in  units with $c = G = 1$), where $\beta_a$ is the ALP velocity, $r$ is the radius where it is produced, and $M(r)$ the mass enclosed in this radius. The spectrum of the gravitationally unbounded ALPs, integrated over a time window of $10\,\s$ after the core bounce, is shown in Fig.~\ref{fig:Spectra} (continuous curves) for two representative masses, $m_a= 20$~MeV (black curve) and $m_a= 200$~MeV (red curve), and for $g_{ap}=10^{-10}$. Since this value is well below the cooling bound for $m_a \ll 1\MeV$ (see next section), the ALPs feedback on the neutrino signal can be reasonably neglected in the whole time domain. As a consequence, we are allowed to integrate the production spectra over the unperturbed SN profiles in the considered time window.
In particular, the curves were obtained numerically, using a 1D spherically symmetric and general relativistic hydrodynamics model of an $18~M_{\odot}$ progenitor, based on the {\tt AGILE BOLTZTRAN} code~\cite{Mezzacappa:1993gn,Liebendoerfer:2002xn}.
The shaded regions in the figure, delimited by the dashed lines, show the spectrum of ALPs with energies $m_a< \omega_a < m_a/\alpha$, which are trapped in the stellar gravitational field. We will discuss some phenomenological implications of these gravitationally bounded ALPs in Sec.~\ref{sec:SN_halo}.
Assuming the pion abundance given in Ref.~\cite{Fore:2019wib}, Fig.~\ref{fig:Spectra} shows that the ALP spectrum corresponding to $m_a=20$~MeV has a bimodal shape, with the lower peak at $\omega_a \sim 80$~MeV due to the $NN$ process, and the higher peak at $\omega_a \sim 200$~MeV associated to the $\pi N$ production. This behavior is not present for the $m_a=200$~MeV case where, because of the mass threshold, $NN$ gives a subdominant contribution included in the pion conversion peak.

\subsection{ALP Luminosity }

The total (energy integrated) luminosity of ALPs produced in the PNS and observed at infinity  can be written as 
\begin{equation}
L_a=    \int d^{3}r \,\ \alpha^2(r) \int d\omega_a    \omega_a \frac{d{n}_{\rm a}}{d {\omega_a}dt} \,\ . 
\label{eq:Lum}
\end{equation}
\begin{figure}[t!]
\centering
	\includegraphics[width=1\columnwidth]{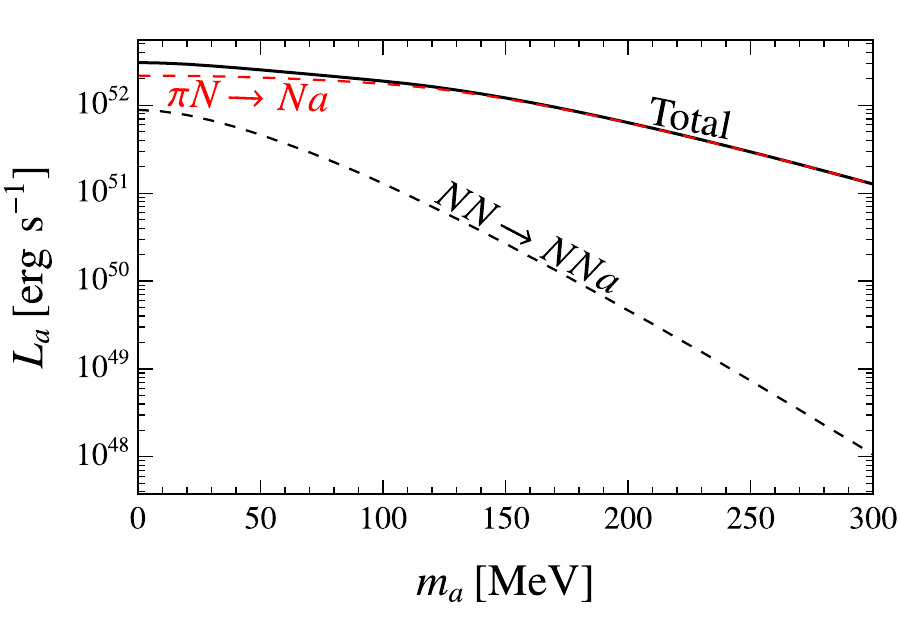}   
	\caption{ALP luminosity at post-bounce time $t_{\rm pb}=1$~s as a function of the ALP mass for $g_{ap}=8.5\times10^{-10}$ for NN Bremsstrahlung (black dashed line), pion conversion $\pi N$ (red dashed line) and total (black solid line).}
	\label{fig:FullLum}
\end{figure}
Figure~\ref{fig:FullLum} shows the ALP luminosities, at the post-bounce time $t_{\rm pb}=1$~s, as a function of the ALP mass for ${g_{ap}=8.5\times10^{-10}}$, which is close to the cooling bound we will find in the massless case.
As discussed before, the pion conversion $\pi N$ (red dashed line) provides the dominant contribution to the total luminosity (black solid line). Furthermore, the $NN$ Bremsstrahlung rate becomes more and more subdominant as the ALP mass increases. This is due to the Boltzmann suppression of the rates for high ALP masses. This suppression is more efficient for Bremsstrahlung ALPs since their average energies are considerably smaller than in the case of the pionic processes~\cite{Carenza:2020cis,Fischer:2021jfm}.\newline In a model-independent way,  it is possible to express the ALP luminosity as a function of the ALP-nucleon couplings through the following fitting formula
\begin{equation}
    L_a\simeq \epsilon_{0}\,(g_{an}^2+b\, g_{ap}^2-c\,g_{an}g_{ap})\,\times 10^{70} \erg\s^{-1}\,,
\label{eq:LaSchematic}
\end{equation}

where $\epsilon_{0}$, $b$, $c$ are fitting parameters. Notice that the minus sign in front of the $c$ parameter is due to the dependence of $C_{a\pi N}$ and $C_{aN \Delta}$ on $C_{ap}$ and $C_{an}$. Table~\ref{tab:FittingParam} gives the values of these parameters for some representative ALP masses.

\subsection{Cooling bound}
\label{subsec:4}

\begin{figure*} [t!]
	\centering
\includegraphics[scale=0.75]{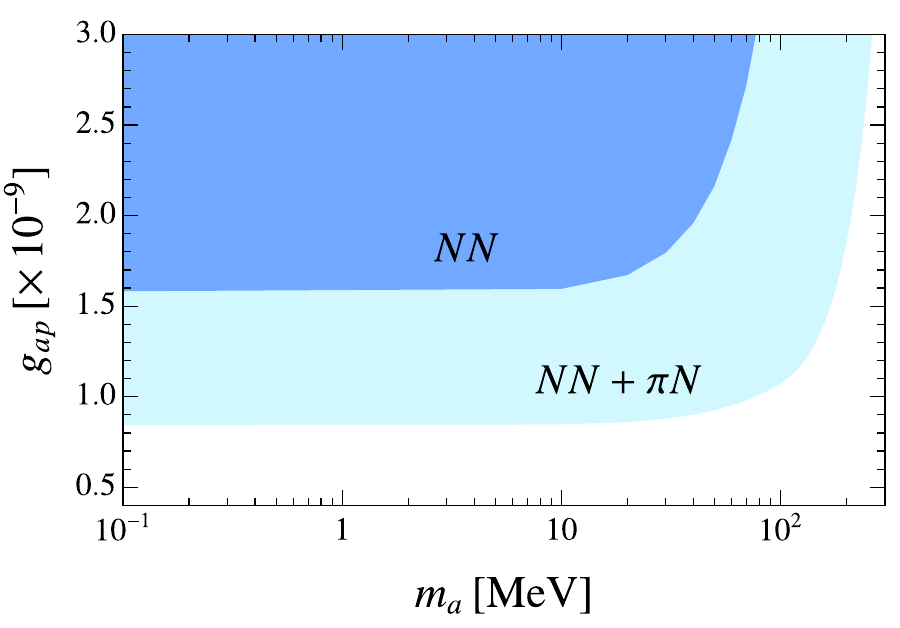}   
	\caption{Exclusion plot in the plane $g_{ap}$ \emph{vs} $m_a$  by means of the SN 1987A cooling. The dark blue region refers to $NN$ Bremsstrahlung only, while the light blue is referred to the total luminosity including also pionic processes.}
	\label{fig:CoolingBound}
\end{figure*}

The luminosity calculated in Eq.~\eqref{eq:Lum} can be used to set a constraint on the ALP-nucleon coupling by imposing at post-bounce time $t_{\rm{pb}}=1\,\s$~\cite{Raffelt:1990yz,Caputo:2021rux} 
\begin{equation}
    L_a\lesssim L_\nu,
\label{CoolingBound}
\end{equation}

to avoid an excessive energy-loss that would shorten the duration of the neutrino burst observed during the SN 1987A explosion. For our SN model \mbox{$L_\nu\simeq3\times 10^{52}\,\erg\,\s^{-1}$} at $t_{\rm{pb}}=1\,\s$. Assuming ALPs coupled only to protons, this constraint excludes the light-blue region in the ALP parameter space shown in Fig.~\ref{fig:CoolingBound}. As expected, the bound obtained by including only the $NN$ Bremsstrahlung (dark-blue region) is significantly less stringent. In particular, for ALP masses $m_{a}\gtrsim m_{\pi}$ the bound is essentially fixed solely by the pion conversion. For $m_a \ll 1\MeV$, the bound becomes $g_{ap}\lesssim8.5\times10^{-10}$, which updates the result of Ref.~\cite{Carenza:2020cis}.
\begin{table}[t!]
 \centering
\begin{tabular}{c c c c}
\hline
$m_a$ (MeV) & $\epsilon_{0}$ &$b$  &$c$\\
\hline
\hline
0 & \,\ \,\  3.86 $ $ $ $  $ $ &  \,\ \,\ 1.10 $ $ $ $  $ $  & \,\ 0.26 $ $ \\
30 &3.32 &1.17 &0.36 \\
90 &2.03 &1.37 &0.79 \\
150 &1.13 &1.48 &1.11 \\
\hline
\end{tabular}
 \caption{Fitting parameters of ALP luminosity  in Eq.~\eqref{eq:LaSchematic} for representative values ALP masses.}
\label{tab:FittingParam}
\end{table}
As the ALP-nucleon coupling increases, the probability of ALP interactions with the nuclear medium gets larger and, eventually, ALPs may no longer be able to freely stream out of the star. This is known as the \emph{trapping regime}. After this regime sets in, the efficiency of the ALP cooling diminishes with increasing couplings. A rough criteria for the trapping regime can be defined imposing that the ALP mean free path~\cite{Giannotti:2005tn}
\begin{equation}
    \lambda_{a}^{-1}=2\pi^2\left(\omega_a^2-m_a^2\right)^{-1}\,\exp{\left(\frac{\omega_a}{T}\right)}\,\frac{dn_a}{d\omega_a\,dt} \,\ ,
\end{equation}
is comparable to the radius of the PNS. 
A crude estimate indicates that this condition is satisfied for \mbox{$g_{ap}\gtrsim2.5\times10^{-9}$}. Trapped ALPs may contribute significantly to the energy transport in the SN core~\cite{Raffelt:1987yt,Caputo:2022rca,Caputo:2021rux}. Thus, complementary bounds can be placed on the ALP-nucleon couplings in this regime. Anyhow, a reliable study of the ALP emission rate in this regime
would require a dedicated numerical analysis that we postpone to a future work.

\section{ALP coupling with photons}
\label{sec:4}

\subsection{ALP radiative decay}

Going beyond the minimal model of ALPs interacting only with nucleons, in the next two Sections we consider ALPs coupled also with photons or leptons (never both at the same time), besides nucleons.
We begin by considering the ALP-photon coupling, described by the following Lagrangian term~\cite{DiLuzio:2020wdo}
\begin{equation}
    {\cal L}_{a\gamma} =-\frac{1}{4}{g_{a\gamma}}  F_{\mu \nu} \tilde F^{\mu\nu} a \, ,
\end{equation}
where $g_{a\gamma}$ is the ALP-photon coupling (in units of inverse energy), $ F_{\mu \nu}$ is the electromagnetic tensor and $\tilde F^{\mu \nu}$ its dual. 

The photon coupling would allow for the Primakoff production of ALPs in a SN core~\cite{Payez:2014xsa}. However, given the relatively strong current constraints on $g_{a\gamma}$ for \mbox{$m_a \gtrsim 1$~MeV} (see, e.g., Ref.~\cite{Lucente:2020whw}, which places the bound $g_{a\gamma}\lesssim5\times10^{-9}\GeV^{-1}$), this production channel would be subleading with respect to the ones associated to the nucleon couplings~\cite{Jaeckel:2017tud} in the region of the parameters space considered. Therefore, we will ignore it in the following discussion. Nevertheless, despite the photon coupling is inefficient in the ALP production in the presence of nucleon coupling, it may open the channel of radiative ALP decay leading to an interesting phenomenology. The ALP radiative decay has a rate~\cite{Raffelt:2006rj}
\begin{equation} \Gamma_{a\gamma\gamma}=g_{a\gamma}^2\frac{m_a^3}{64\pi}\,.
\label{eq:decayrate}
\end{equation}
This rate corresponds to a decay length~\cite{Calore:2021klc} 
\begin{equation}
\begin{split}
        \lambda_\gamma&=\frac{\gamma_a \beta_a}{\Gamma_{a\gamma\gamma}}\simeq 1.3\kpc\left(\frac{\omega_{a}}{100\MeV}\right)\left(\frac{m_{a}}{10\MeV}\right)^{-4}\\
        &\times\left(\frac{g_{a\gamma}}{10^{-13}\GeV^{-1}}\right)^{-2}\sqrt{1-\left(\frac{\omega_a}{m_a}\right)^2}\,,
    \end{split}
\end{equation}
where $\gamma_a$ is the ALP Lorentz factor. 
Depending on the value of $\lambda_\gamma$, ALPs can decay inside or outside the SN, giving a qualitatively  different phenomenology. 
In particular, taking for definiteness the envelope radius of type II SNe $R_{\rm env}\sim3\times10^{14}~\cm$~\cite{Calore:2021lih}, for $\lambda_\gamma < R_{\rm env}$, ALP decays can deposit energy inside the SN, while for $\lambda_\gamma > R_{\rm env}$ they decay outside the star, producing an observable gamma-ray signal. 
We will discuss both these effects in the following Sections. 
For the sake of clarity, in the following analysis we fix the axion-proton coupling to $g_{ap}=10^{-10}$, a value sufficiently small to ignore the ALP feedback on the cooling mechanism and integrate the spectra over the unperturbed profiles.

\subsection{ALP energy deposition inside the SN envelope}
\label{ALP energy deposition}

\begin{figure*} [t!]
	\centering
\includegraphics[scale=0.75]{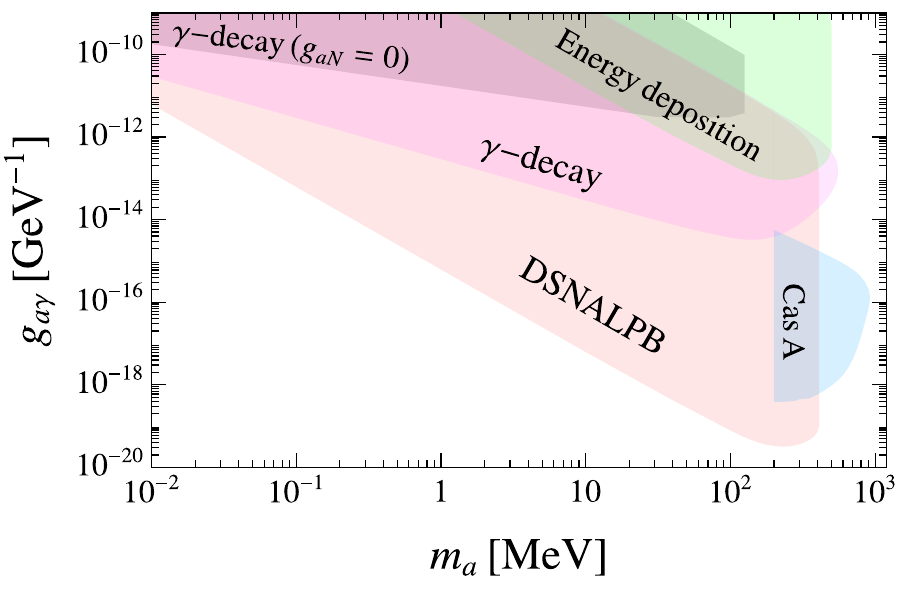}
	\caption{Bounds on the ALP-photon coupling $g_{a\gamma}$ for massive ALPs from SNe at $g_{ap}=10^{-10}$ and $g_{an}=0$. The green area is referred to the energy deposition mechanism in SN 1987A, the magenta region is associated to radiative decays of escaping ALPs from SN 1987A, the pink area is the excluded region by means of the DSNALPB while the light blue one refers to the decay of ALPs trapped inside the Cas A gravitational potential. Finally, the gray region is the excluded region discussed in Ref.~\cite{Jaeckel:2017tud}, considering $\gamma$-decays of massive ALPs produced by Primakoff processes (see text for more details).
	} 
	\label{fig:PhotonBound}
\end{figure*}

We consider at first the case in which ALPs decay inside the SN.  Following the analysis in Ref.~\cite{Caputo:2022mah}, massive ALPs decaying into photons can deposit energy in the outer layers of the SN volume. In particular, if the decay length satisfies $R_{\rm PNS}<\lambda_\gamma<R_{\rm env}$, the whole energy produced in nuclear processes in the form of decaying ALPs is dumped into the progenitor star. This deposited energy can contribute to the SN explosion, providing a  ``calorimetric'' constraint to radiative decays~\cite{Falk:1978kf,Sung:2019xie}. At this regard, Ref.~\cite{Caputo:2022mah} used a SN population with particularly low explosion energies as the most sensitive calorimeters to constrain this possibility. Their low energies limit the energy deposition from particle decays to less than about $0.1$~B, where $1$ B (Bethe) $=10^{51}$~erg. The energy dumped inside the mantle can be computed as~\cite{Caputo:2022mah} 
\begin{eqnarray}
     E_{\rm dep}=\int dt \int_0^{R_{\rm PNS}} dr 4\pi r^2 \int_{0}^\infty d\omega_a \omega_a \frac{dn_ a}{d {\omega_a}\,dt}   \nonumber\\
     \times\left[ e^{-(R_{\rm PNS}-r)/\lambda_\gamma}\,-e^{-(R_{\rm env}-r)/\lambda_\gamma}\right]\,,
\label{eq:Emantle}
\end{eqnarray}
where the radial integral extends over the ALP production region. The first exponential in the r.h.s. implies that only ALP decaying at $r> R_{\rm PNS}$ contribute to the energy deposition, while the second exponential selects only photons produced in decays within $R_{\rm env}$. Requiring that the deposited energy in the mantle is less than $0.1$~B, we have obtained the bounds in the plane $g_{a\gamma}$ \emph{vs} $m_a$ shown in the green region in Fig.~\ref{fig:PhotonBound}. Notice that the constraint extends only to $m_a\lesssim450~\MeV$. For heavier ALPs, the bound disappears since the Boltzmann suppression is so large that the energy deposited in the mantle is less than $0.1$~B, even if all the produced ALPs decay inside it.

\subsection{Non-observation of gamma-rays from SN 1987A}
If the ALP decay length $\lambda_\gamma$ is sufficiently large, ALPs would escape the SN envelope and if they decay before reaching the Earth, they would originate an observable photon signal. Following the analysis displayed in Ref.~\cite{Caputo:2021rux}, we take as a reference the observation performed by the Gamma-Ray Spectrometer (GRS) on board of the Solar Maximum Mission (SMM), working during the SN 1987A explosion event, occurred at a distance $d_{\rm SN}=51.4$~kpc. Since no excess in photon counts in the energy range $25-100 \MeV$ was observed in the time interval $T=232.2 \s$ after the detection of the neutrino burst, the induced photon flux must be smaller than the $\gamma$-fluence limit of the detector~\cite{OBERAUER1993377}
\begin{equation}
    \phi_\gamma\lesssim1.38\,\cm^{-2}\,.
\end{equation}
The differential photon flux expected at the Earth as a consequence of the decay of the emitted ALPs can be computed analogously to Eq.~(51) of Ref.~\cite{Caputo:2021rux} and Eq.~(18) of Ref.~\cite{OBERAUER1993377} as 
\begin{equation}
\begin{split}
    \frac{d\phi_\gamma}{dE_\gamma dt}=& \frac{2}{4\pi d_{\rm SN}^2} \,\ e^{-\frac{R_{\text{env}}}{\lambda_\gamma}}\frac{2\,E_\gamma\Gamma_{a\gamma\gamma}}{m_a}e^{-\frac{2\,E_\gamma\Gamma_{a\gamma\gamma}}{m_a}t}\\
    &\times \int_{E_\gamma}^\infty d\omega_a p_a^{-1}\, \frac{dN_a}{d\omega_a}\,,
\end{split}
\label{eq:decayflux}
\end{equation}
which is derived extending the discussion about the neutrino decays in Ref.~\cite{OBERAUER1993377,Jaffe:1995sw}. Here, $\Gamma_{a\gamma\gamma}$ is the rest-frame ALP decay rate given by Eq.~\eqref{eq:decayrate}, while the integral term is obtained considering a parent particle decaying into photons with energy uniformly distributed in the interval $\left[(\omega_a-p_a)/2,(\omega_a+p_a)/2\right]$, with $p_a=\sqrt{\omega_a^2-m_a^2}$ [see Eq.~(16) of Ref.~\cite{OBERAUER1993377} for further details]. Moreover we have introduced the exponential factor $\exp{[-R_{\rm env}/\lambda_\gamma]}$ to exclude from the analysis ALPs which decay inside the SN volume.

The resulting constraint on the ALP parameter space is shown in the magenta region reported in Fig.~\ref{fig:PhotonBound}. Remarkably, our bound significantly enlarges the excluded region, obtained in Ref.~\cite{Jaeckel:2017tud}, assuming the Primakoff process as the only ALP production channel (the gray area in Fig.~\ref{fig:PhotonBound}). We mention that the upper limit of our bound reaches lower values of $g_{a\gamma}$ since in Ref.~\cite{Jaeckel:2017tud} an envelope radius of $3\times 10^{12}$~cm was assumed, two orders of magnitude lower than the one used here.

\subsection{Diffuse SN ALP background}
\label{susec:6} 
In this Section we are going to revisit another enthralling phenomenolgical prediction of ALPs coupled with nucleons and photons: the existence of a Diffuse SN ALP Background (DSNALPB)~\cite{Raffelt:2011ft},  analogous to the diffuse neutrino background~\cite{Beacom:2010kk}, generated by the ALP emission from all SN explosions occurred at an epoch recent enough to be contained in the current event horizon. The radiative decay of the DSNALP has been studied in Ref.~\cite{Eckner:2021kjb,Calore:2021hhn}. The results presented in this work allow us to revisit the previous analyses by including the corrections to the Bremsstrahlung production rate in the SN core induced by a finite ALP mass. Moreover here we include also the contribution due to pionic processes.

To obtain the diffuse photon flux induced by decays of the DSNALPB it is necessary to integrate above the redshift to take into accounts all the contributions coming from the past core-collapse SNe in the Universe as done in Ref.~\cite{Calore:2020tjw}
\begin{equation}
    \frac{d\phi_\gamma^{\rm dif}}{dE_\gamma}=\int_0^\infty(1+z)\frac{dN_\gamma(E_\gamma(1+z))}{dE_\gamma} [R_{SN}(z)]\left[\Bigg|\frac{dt}{dz}\Bigg|dz\right]\,.
\label{eq:GammaFluxDSNALPB}
\end{equation}
In this expression $R_{SN}(z)$ is the SN explosion rate, taken from Ref.~\cite{Priya:2017bmm}, with a total normalization for the core-collapse rate $R_{cc}=1.25\times10^{-4}\text{yr}^{-1}\Mpc^{-3}$ and computed assuming an average progenitor mass of about $18$~$M_{\odot}$, as the one considered until now. Furthermore, here $|dt/dz|^{-1}=H_0(1+z)[\Omega_\Lambda+\Omega_M(1+z)^3]^{\frac{1}{2}}$ with the cosmological parameters $H_0=67.4 \km \s^{-1} \Mpc^{-1}$, $\Omega_\Lambda=0.7$ and $\Omega_M=0.3$. Moreover the photon spectrum $dN_\gamma/dE_\gamma$ induced by the decay of heavy ALPs, observed at a distance $d(z)$ from the ALP source can be computed as~\cite{Calore:2020tjw}
\begin{equation}
    \frac{dN_\gamma(E_\gamma)}{dE_\gamma}=2\times\left[1-e^{-d/\lambda_\gamma}\right] \frac{dN_a(2E_\gamma)}{d\omega_a}e^{-R_{\text{env}}/\lambda_\gamma}\,,
\end{equation}
where the last exponential factor takes into account a possible fraction of ALPs which decay inside the SN envelope and does not contribute to the DSNALPB. Indeed, we notice that for ALP masses $m_a>100\MeV$ the decay lengths are comparable to $R_{\rm env}$. Following Ref.~\cite{Calore:2020tjw}, in order to constrain the gamma-ray flux associated with DSNALPB decays we use the isotropic $\gamma$-ray background measurements provided by the \textit{Fermi}-LAT Collaboration, by means of  Pass 8 R3 processed data (8-yr dataset) for the ULTRACLEANVETO event class section. For $E\geq50 \MeV$ this data can be well fitted with a power law~\cite{Calore:2020tjw}
\begin{equation}
    \frac{d\phi_\gamma(E)}{dE_\gamma}=2.2\times10^{-3}E^{-2.2} \MeV^{-1} \cm^{-1} \s^{-1} \text{sr}^{-1}\,.
\label{eq:FermiData}
\end{equation}

\begin{figure}[t!]
	\centering
\includegraphics[width=1\columnwidth]{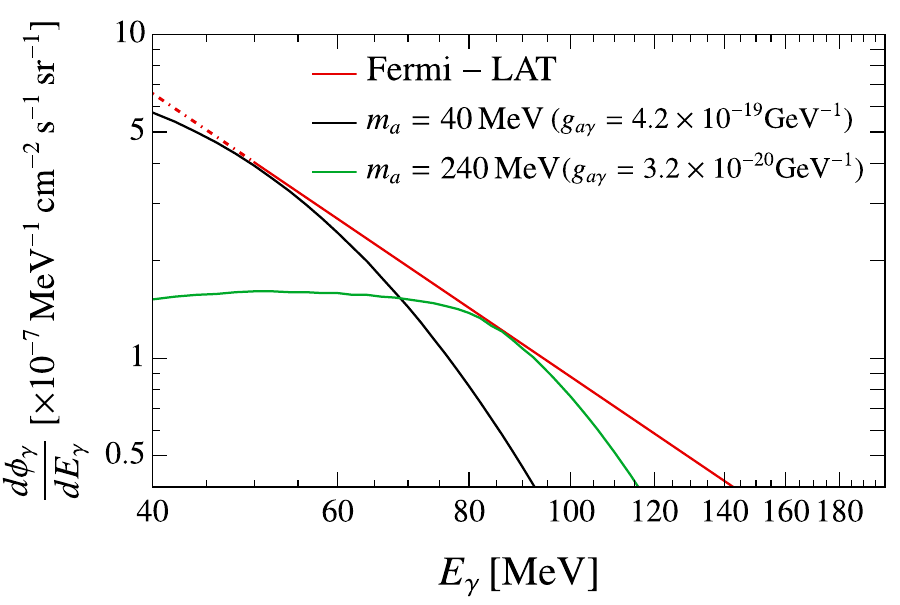} 
    \caption{Diffuse gamma-ray fluxes induced by the decay of DNSALPB assuming ALPs with given masses. The red line refers to the diffuse gamma-ray background measured by {\it Fermi}-LAT. The dot-dashed part of the line is an extrapolation of the Fermi line at energies $30\MeV\leq E \leq 50\MeV$.}
	\label{fig:DSNALPBfluxes}
\end{figure}

In Fig.~\ref{fig:DSNALPBfluxes} we compare the diffuse gamma-ray flux from DSNALPB decays for different choices of $g_{ap}$ and $m_a$ with the gamma-ray flux measured by \textit{Fermi}-LAT. We exclude parts of the ALP parameter space that would give a gamma-ray flux exceeding the measured one. Our excluded region is reported as the pink one in the ALP parameter space shown in Fig.~\ref{fig:PhotonBound}. In particular it extends the results reported in Ref.~\cite{Calore:2020tjw} to ALP masses up to $m_a\simeq400 \MeV$, while larger masses cannot be probed due to an inefficient ALP production.

\subsection{Gamma-ray halo from SN remnants}
\label{sec:SN_halo}

As discussed in Sec.~\ref{subsec:ALPspectrum}, a portion of the heavy ALPs produced in the SN core would not have sufficient kinetic energy to escape its strong gravitational potential. These ALPs would accumulate around the SN remnant forming an  ``ALP halo''. To gain some intuition about the properties of this halo, let us assume that ALPs are generated at a typical radius $R\approx 10\,$km, and let us model the lapse factor as in the Schwarzschild case, $\alpha=\left( 1-\frac{2GM}{R} \right)^{1/2}$. Working in the Newtonian limit (which here is a valid assumption, since $\alpha\approx 1$, as evident from Eq.~\eqref{eq:lapse_factor_weak_limit}), the condition for gravitational trapping, $\omega_a\leq m_a/\alpha$, leads to a relation for the maximal radius of the cloud. Exploiting the conservation of energy $K-GMm_a/R=-GMm_a/R_{\rm max}$ one obtains
\begin{align}
\label{eq:lapse_factor_weak_limit}
\frac{R_{\rm max}}{R}=\left(1-\frac{K R}{GMm_a}\right)^{-1}\,,
\end{align}
where $K$ is the ALP kinetic energy. 
Notice that  

\begin{equation}
\label{eq:}
\begin{split}
\frac{K R}{GMm_a}\approx 0.24 
&\left(\frac{R}{10\,{\rm km}}\right)
\left(\frac{1.4\,M_{\odot}}{M}\right)\times\\
&\left(\frac{K}{10\,{\rm MeV}}\right)
\left(\frac{200\,{\rm MeV}}{m_a}\right)\,.
\end{split}
\end{equation}

Thus, for ALPs with mass $200$~MeV, the maximum kinetic energy permitted by the trapping condition is $K_{\rm max}\approx 40$~MeV. This is in good agreement with the numerical analysis shown in Fig.~\ref{fig:Spectra}, where we see that the unbounded spectrum of the $200$~MeV ALPs (continuous red line) begins at about $240$~MeV.   Therefore, the SN ALP halo is composed of slow (non-relativstic) ALPs, with kinetic energies of a few MeV.

If ALPs couple to photons, their decay could lead to a detectable MeV photon signal from the ALP halo. Notice that a similar argument was proposed in Ref.~\cite{Hannestad:2001xi} for the case of massive Kaluza-Klein gravitons. The signal would appear as point source in  gamma-ray telescopes. In fact, according to Eq.~\eqref{eq:lapse_factor_weak_limit}, the radius of the cloud is expected to be quite small, on astronomical scales, except for ALPs with energies extremely close to the maximal allowed kinetic energy. For example, ALPs with energy within $1\%$ of the maximal kinetic energy might be found no further than $\sim 10^3$ km from the SN remnant. 

According to Ref.~\cite{Calore:2020tjw}, the photon flux per unit energy expected on Earth after a time $t$ from the moment the explosion was observed, can be computed as 
\begin{equation}
        \frac{d\phi_\gamma^{\rm halo}}{dE_\gamma}=\frac{2}{4 \pi d^2}\left(\frac{dN_a(2 E_\gamma)}{dE_\gamma}\right)_{\rm{trap}}\,\frac{\Gamma_{a\gamma\gamma}}{\gamma_a}e^{-\frac{\Gamma_{a\gamma\gamma} t}{\gamma_a}}\,,
\label{eq:photonFluxEx}
\end{equation}
where $d$ is the Earth-SN distance, $E_\gamma$ is the energy of the emitted photon and $\gamma_a$ is the Lorentz factor of the ALP with energy $2\,E_\gamma$. Here, $\Gamma_{a\gamma\gamma}/\gamma_a\exp{[-\Gamma_{a\gamma}/\gamma_a\,t]}$ is the probability per unit time to observe an ALP decay, while the spectrum of the gravitationally trapped ALPs could be simply computed as
\begin{equation}
    \left(\frac{dN_a}{d\omega_a}\right)_{\rm trap}=\frac{dN_a}{d\omega_a}\Bigg|_{\alpha=1}-\,\,\frac{dN_a}{d\omega_a}\Bigg|_{\alpha\neq1}\,,
\end{equation}
where, as explained in Sec.~\ref{subsec:ALPspectrum},  $dN_a/d\omega_a|_{\alpha=1}$ and $dN_a/d\omega_a|_{\alpha\neq1}$ are the ALP spectra with and without gravitational effects, respectively.

A favorable environment to apply our considerations is Cassiopeia A (Cas A), which is a supernova remnant in the constellation of Cassiopeia at a distance $d\simeq3.4 \kpc$. The explosion occurred  $t\simeq320\,\rm{yrs}$ ago. Following Ref.~\cite{Hannestad:2001xi}, since the Fermi-LAT experiment has not found any gamma-ray signal in the range $100\MeV$-$1 \GeV$ close to the site of Cas A~\cite{Yuan:2013qwa}, one can put the following limit on the photon flux for photon energies in this range
\begin{equation}
    \phi_{E>100 \MeV}\lesssim 2\times10^{-8} \cm^{-2} \s^{-1}\,,
\label{PhotonFluxConst}
\end{equation}
where this value is the sensitivity of the Fermi-LAT experiment in this range of energies. Applying this constraint on the photon flux computed as in Eq.~(\ref{eq:photonFluxEx}), which depends both on $g_{aN}$ and $g_{a\gamma}$, we can introduce the bound depicted as the light blue area in Fig.~\ref{fig:PhotonBound} for ALP masses $m_a\geq200 \MeV$.

\section{\label{sec:5} ALP coupling with leptons}
\subsection{ALPs leptonic decays}

As next case we can consider the ALP coupling with leptons, described by the following Lagrangian~\cite{DiLuzio:2020wdo}
\begin{equation}
    {\cal L}_{a \ell} =\sum_{\ell=e,\mu}\frac{g_{a\ell}}{2m_\ell}(\Bar{\ell}\gamma_\mu\gamma_5\ell)\partial^\mu a\,,
\end{equation}
where the sum runs over the lepton flavors (electron and muon) and $g_{a\ell}$ is the ALP coupling to the lepton $\ell$ with mass $m_\ell$. 
In a SN, ALPs coupled with electrons would be produced via electron-proton Bremsstrahlung \mbox{($e^- \, p \to e^-\,p\,\,a$)} and electron-positron fusion \mbox{($e^-\,e^+ \to a$)}~\cite{Lucente:2021hbp,Ferreira:2022xlw}, while the dominant contributions due to muons are the Compton scattering ($\gamma\,\mu \to a\,\mu$) and the muon-proton Bremsstrahlung \mbox{($\mu\,p\to \mu\,p\,a$)}~\cite{Bollig:2020xdr,Croon:2020lrf}. However, in almost the entire range of parameters we are exploring in this work, the contribution of the lepton processes to the SN ALP production remains subdominant with respect to the nuclear production mechanisms.
A possible exception will be discussed in Sec.~\ref{sec:gaeenv}.

Due to their couplings to leptons, massive ALPs can decay into leptons trough pair productions $a\to\Bar{\ell}\,\ell$ if $m_a>2m_\ell$. The decay length for these processes is given by~\cite{Jaeckel:2017tud,Altmann:1995bw}: 
\begin{align}
\lambda_{\ell}&=\frac{8\pi}{g_{a\ell}^2m_a}\frac{\omega_a}{m_a}
\sqrt{\frac{1-m_a^2/\omega_a^2}{1-4m_\ell^2/m_a^2}} =  \nonumber\\ 
&=0.16\,{\rm kpc}\left( \frac{g_{a\ell}}{10^{-15}} \right)^{-2}  \left( \frac{m_a}{1\,{\rm MeV}} \right)^{-1}\,\sqrt{\frac{\omega_a^2-m_a^2}{m_a^2-4m_\ell^2}}\,.
\end{align}
For the sake of the simplicity, in the following we consider ALPs coupled with a single species of leptons (electrons or muons) at a time. We remark that the ALP-lepton coupling induces at one loop an ALP photon coupling, which for masses $m_a>2\,m_\ell$ is given by~\cite{Craig:2018kne,Caputo:2021rux,Ferreira:2022xlw,Ferreira:2022egk}
\begin{equation}
    g_{a\gamma}^{\rm eff} = \frac{\alpha}{\pi\,m_\ell}\,g_{a\ell}\,,
    \label{eq:gageff}
\end{equation}
with $\alpha=1/137$. Due to this induced coupling, ALPs interacting with electrons may decay into photons. This allowed the authors of Ref.~\cite{Ferreira:2022xlw} to constrain $g_{ae}$ from the non-observation of gamma-rays after the SN 1987A explosion. 
However, it is easy to check that in the mass range we are studying in this work ($m_a\lesssim 300~\MeV$) lepton decays are dominant over the induced photon channel. Therefore, we can neglect this 
subleading effect hereafter.

\subsection{ALP energy deposition inside the SN envelope}
\label{sec:gaeenv}

\begin{figure} [t!]
	\centering
\includegraphics[width=1\columnwidth]{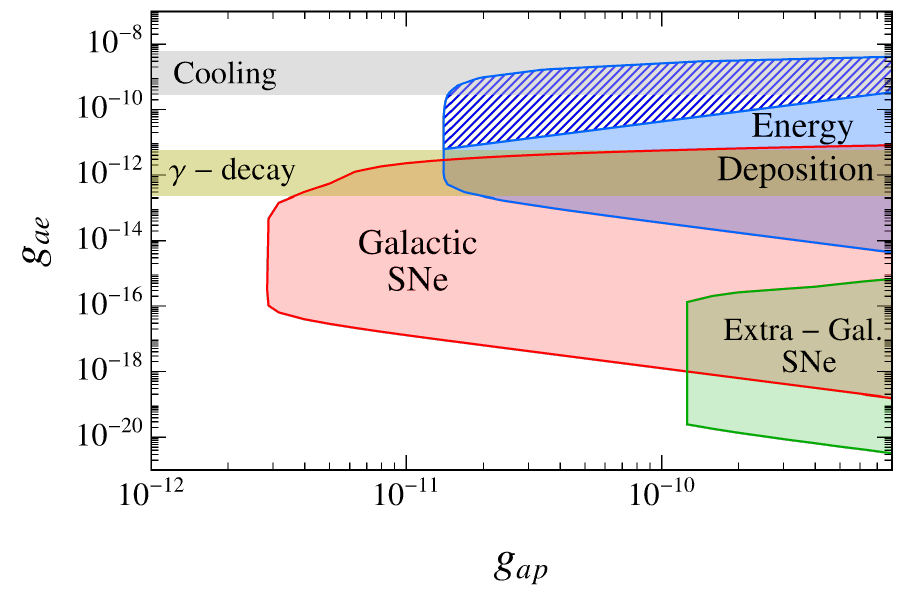}
	\caption{Bounds in the plane $g_{ae}$ {\emph vs} $g_{ap}$ for an ALP mass $m_a=100 \MeV$. In the blue hatched region, the energy deposition bound is uncertain since ALP production by means of electron-positron fusion becomes competitive to nuclear processes. Furthermore, we show in gray the cooling bound~\cite{Ferreira:2022xlw} and in dark yellow the loop-induced radiative decay bound on $g_{ae}$ discussed in Ref.~\cite{Ferreira:2022xlw}.
	}
	\label{fig:gaegap100}
\end{figure}

\begin{figure*} [t!]
\centering
\includegraphics[scale=0.8]{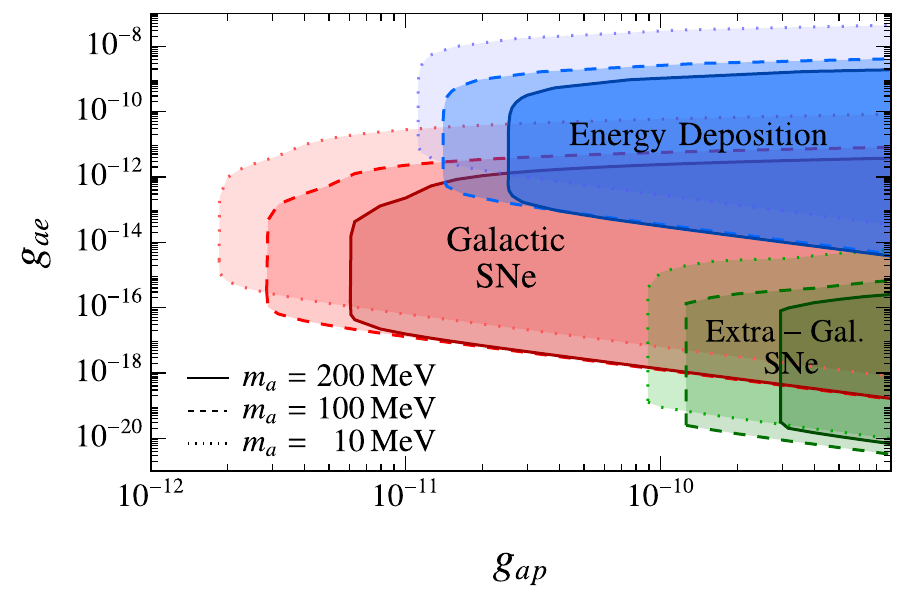}
\includegraphics[scale=0.8]{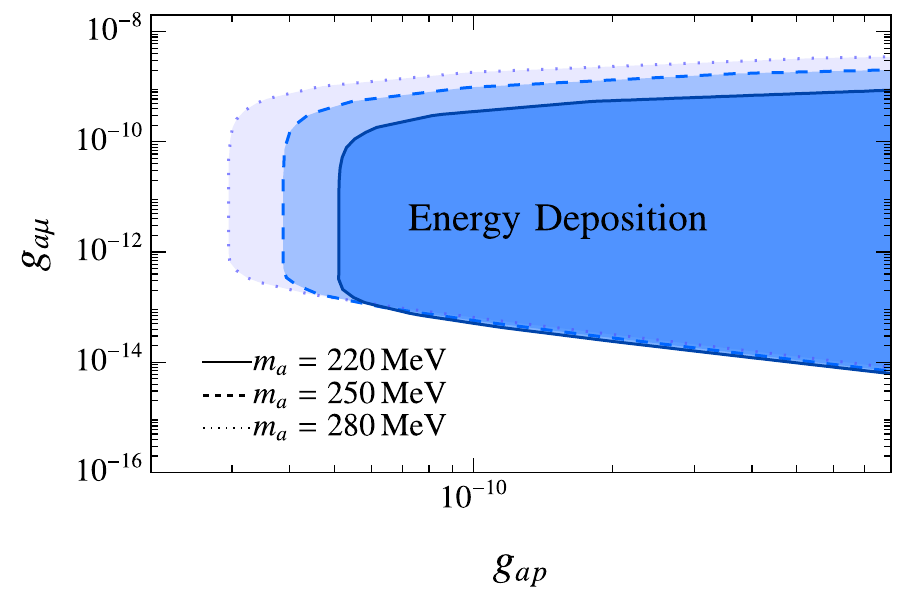}
\caption{\emph{Upper panel}: Excluded regions in the plane $g_{ap}${\emph vs} $g_{ae}$ due to energy deposition (blue regions), 511 keV line from Galactic  (red regions) and Extra-galactic (green regions) SNe, for three different values of the ALP mass $m_a=10,~100,~200$~MeV. \emph{Lower panel}: Excluded regions from energy-deposition bounds in the plane $g_{ap}$ {\emph vs} $g_{a\mu}$, for three different values of the ALP mass $m_a=220,~250,~280$~MeV.}
	\label{fig:galgap}
\end{figure*}

ALPs decaying into charged leptons inside the SN envelope would deposit energy in the form of $e^+\,e^-$ or $\mu^+\,\mu^-$, contributing to the SN explosion energy. Proceeding as in the photon decay case, presented in Sec.~\ref{ALP energy deposition}, we compute the energy dumped in the SN mantle by leptonic decays substituting $\lambda_\gamma$ with $\lambda_\ell$ in Eq.~\eqref{eq:Emantle} and obtain the corresponding constraints. In particular, in Fig.~\ref{fig:gaegap100} we consider the plane $g_{ae}$ \emph{vs} $g_{ap}$ at fixed value of the ALP mass $m_a = 100$~MeV. As shown by the blue region, at $g_{ap}\approx8\times 10^{-10}$ we exclude values of the ALP-electron coupling $4\times10^{-15}\lesssim g_{ae} \lesssim 4\times 10^{-9}$. As $g_{ap}$ decreases, the excluded region becomes smaller until it disappears completely for $g_{ap}\lesssim 1.5\times 10^{-11}$, due to the inefficient ALP production. The horizontal gray band is excluded by the cooling bound on the ALP-electron coupling~\cite{Lucente:2021hbp,Ferreira:2022xlw}, while the yellow region is the loop-induced photon decay bound recently discussed in Ref.~\cite{Ferreira:2022xlw}. We notice that for ALP-electron coupling $g_{ae}\gtrsim 10^{-10}$ the ALP production via electron processes becomes significant and might be competitive with nuclear processes (see the hatched region in Fig.~\ref{fig:gaegap100})~\cite{Lucente:2021hbp,Ferreira:2022xlw}.

To analyze the mass dependence of the bound, in the upper panel of Fig.~\ref{fig:galgap} we show the constraints in the $g_{ae}$ {\emph vs} $g_{ap}$ plane for three different values of the ALP mass $m_a = 10$ MeV (dotted contours), $m_a = 100$ MeV (dashed contours) and $m_a = 200$ MeV (solid contours). As the mass increases, the energy deposition bound (the blue regions) requires larger values of $g_{ap}$, since the ALP production is Boltzmann suppressed. In addition, it is shifted toward lower values of $g_{ae}$ beacause of the dependence of the ALP decay length on the mass. 
Finally, when the ALP mass becomes larger than the pion mass, the excluded region becomes smaller. The reason is that the ALP production in the pionic Compton process, which is the dominant one at high mass, is kinematically suppressed for masses above the pion mass. Indeed, for kinematical consistency, the ALP energy in the pionic process must be larger than the pion mass.

Finally, ALPs with masses larger than $2\,m_\mu$ are allowed to decay also in $\mu^+\mu^-$. Therefore, the ALP-muon coupling can be constrained by the energy deposition argument. In the lower panel of Fig.~\ref{fig:galgap} we show the bounds for $m_a=220,\,250,\,280$~MeV. The shape of the bound is analogous to the electronic decays. In particular, since $m_a > m_\pi$, in this case the excluded region becomes smaller and smaller as the ALP mass increases, as previously discussed in this Section.

\subsection{ALP decays outside the SN envelope}

As a final possibility, we consider now the case in which ALPs are so weakly coupled to leptons that they  decay outside the SN envelope. If they decay into electron-positron pairs, they would contribute to the Galactic 511 keV photon line\footnote{At the moment, we do not have the knowledge of any observable which can be used to constrain ALPs decaying into $\mu^+\,\mu^-$ outside the SN envelope.}  (see Ref.~\cite{Guessoum:2005cb} for more details on the production mechanism of this line). Using the 511 keV line observations by the spectrometer SPI (SPectrometer on INTEGRAL)~\cite{Calore:2021klc} feebly interacting particles having decay channels with positrons in the final state can be constrained~\cite{Calore:2021lih,Calore:2021klc,DeRocco:2019njg}.

Using the morphology of the signal observed by SPI~\cite{Siegert:2019tus}, in Ref.~\cite{Calore:2021klc} some of us constrained ALPs with mass $m_a\lesssim 30$~MeV simultaneously coupled with nucleons and electrons, produced via nucleon Bremsstrahlung and decaying into electron-positron pairs. Note that ALPs with mass $m_a\gtrsim 30$~MeV were not considered in that paper since their production flux was not characterized yet.
 
In this Section we present the results of a refined analysis of the 511 keV line bound, which accounts for the ALP mass effects in the SN production, allowing us to extend the constraint to heavier ALPs. We recall that the SPI observations imply a bound on the number of positrons injected in the Galaxy~\cite{Calore:2021lih}
\begin{equation}
    N_{\rm pos}\lesssim 1.4\times 10^{52}\,,
    \label{eq:nposbound}
\end{equation}
where $N_{\rm pos}$ is given by
  \begin{eqnarray}
    N_{\rm pos}\!\!& = &\!\!\int d\omega\frac{dN_a}{d\omega}\left(\epsilon_{II}e^{-{R_{\rm env}^{\rm II}}/{\lambda_{e}}}+\epsilon_{I}\,e^{-{R_{\rm env}^{\rm I}}/{\lambda_{e}}}\right) \nonumber
    \\
    &\times&\bigg[1 - \exp\bigg(-\frac{r_{\rm G}}{\lambda_{e}}\bigg) \bigg] \,.
    \label{eq:npos}
\end{eqnarray} 
Following Ref.~\cite{DeRocco:2019njg}, we consider the contribution of two types of core-collapse SNe, namely type Ib/c and type II SNe, characterized by envelope radii $R_{\rm env}^{\rm I}=2\times 10^{12}~{\rm cm}$ and $R_{\rm env}^{\rm II}=10^{14}~{\rm cm}$~\cite{DeRocco:2019njg}, and average fraction $\epsilon_{I}=0.33$ and $\epsilon_{II}=1-\epsilon_{I}$~\cite{Li:2010kd}, respectively. Moreover, $r_{\rm G}=1~\kpc$ is a typical escape radius from the Galaxy.

The red region in Fig.~\ref{fig:gaegap100} represents the Galactic bound from the 511 keV line, obtained by saturating Eq.~\eqref{eq:nposbound}, for $m_a=100$~MeV. Here, we exclude \mbox{$1.5\times 10^{-19}\lesssim g_{ae} \lesssim 8\times 10^{-12}$} at \mbox{$g_{ap}\approx8\times 10^{-10}$} and \mbox{$10^{-16}\lesssim g_{ae} \lesssim 6\times 10^{-14}$} at \mbox{$g_{ap}\approx3\times 10^{-12}$}. As shown by the red regions in Fig.~\ref{fig:galgap} (upper panel), the dependence of the bound on the ALP mass is similar to the energy deposition bound, since the constraint is shifted toward lower couplings as the mass increases and the excluded region becomes smaller for ALP masses larger than the pion mass. We mention that the bound at $m_a= 10~\MeV$ (the light red region) excludes larger values of $g_{ae}$ than the one in our previous work~\cite{Calore:2021klc}, since here we are considering the contribution of both type Ib/c and II SNe, while in Ref.~\cite{Calore:2021klc} we considered only type II SNe, with an escape radius $R_{\rm env}^{\rm II}\gg R_{\rm env}^{\rm I}$. In addition, here the bound reaches lower values of the ALP-nucleon coupling ($g_{ap}\approx 3\times 10^{-12}$) since the ALP production is enhanced by the pion conversion. 

If ALPs are produced by extra-galactic SNe and decay into electron-positron pairs outside our Galaxy, the produced positrons would slow down and annihilate at rest, giving two photons, each one with energy $m_e$, as in the Galactic case. However, these photons are redshifted and thus they contribute to the cosmic X-ray background (CXB), measured by different experiments, e.g. the High Energy Astronomy Observatory (HEAO)~\cite{McHardy:1997fb} and the Solar Maximum Mission (SMM)~\cite{doi:10.1063/1.53933}. In this case, as discussed in Ref.~\cite{Calore:2021klc}, the cumulative energy flux of ALPs from past core-collapse SNe and decaying into electron-positron pairs in a small redshift interval $[z_d,\,z_d-dz_d]$ comes from $z\lesssim 2$ and it is well approximated by~\cite{Calore:2021klc}
\begin{equation}
\begin{split}
 &\bigg(\frac{d^2 \phi_a (E_a)}{d E_a d z_d}\bigg)_{\rm dec}=
 \int_{z_d}^{\infty} \! (1+z) \frac{dN_a(E_a(1+z))}{dE_a} 
\\
& \quad\times
[R_{SN}(z)] \exp\bigg(-\frac{z-z_d}{H_0 \lambda_e} \bigg) \frac{1}{H_0 \lambda_{e}} \bigg[ \bigg|\frac{dt}{dz} \bigg| dz \bigg] \,.
\label{eq:diffuse}
\end{split}
\end{equation}
Due to the redshift, the produced photons reach us with an energy $E_\gamma = m_e/(1+z_d)$, thus the produced photon flux is given by
\begin{eqnarray}
\frac{d \phi_\gamma}{d E_\gamma}
&=&2 k_{ps} \frac{d \phi_a}{d z_d} 
\frac{d z_d}{d E_\gamma}= \nonumber \\
&=&2 k_{ps}
\frac{m_e}{E_\gamma^2}\int_{m_a}^{\infty} dE_a  \bigg(\frac{d^2 \phi_a (E_a)}{d E_a d z_d}\bigg)_{\rm dec} \,\,,
\end{eqnarray}
where $k_{ps}=1/4$ accounts for the fraction of positrons annihilating through parapositronium, producing two photons with energy $m_e$. In this case, we can constrain the ALP parameter space by requiring that the produced photon flux from ALP decays does not exceed the measured CXB by more than 2$\sigma$ (see, e.g., Fig. 8 in Ref.~\cite{Calore:2021klc}). The green region in Fig.~\ref{fig:gaegap100} represents the extra-galactic bound for $m_a=100$~MeV, excluding $ 3\times 10^{-21}\lesssim g_{ae}\lesssim 7\times 10^{-16}$ at $g_{ap}\approx 8\times 10^{-10}$. In this case, $g_{ap}$ can be reduced by less than one order of magnitude before the disappearance of the bound. Also for the extra-galactic bound, as shown by the green regions in the upper panel of Fig.~\ref{fig:galgap}, the mass dependence is analogous to the cases previously discussed.

\section{Conclusions}
\label{sec:conclusions}

In this work, we have presented a detailed study of the PNS emission of heavy ALPs with $m_a \sim {\mathcal O} (100)$~MeV coupled to nucleons, and presented several phenomenological consequences not discussed previously. 

Concerning the emission process, we have extended the previous analysis of the Bremsstrahlung emission rate~\cite{Giannotti:2005tn} beyond the OPE approximation and added the contribution of the Compton pionic process. As we have shown, this last process, which was largely overlooked in the past, dominates the rate in a large section of the massive ALP parameter space. Equipped with this new results, we have been able to extend the SN 1987A cooling bound up to $m_a \sim 300$~MeV and couplings $g_{ap}\gtrsim 8.5\times 10^{-10}$ for the massless case, a limit that extends well below the current experimental constraints and sensitivity.

Though out of reach for direct detection experiments, heavy ALPs with couplings in the region allowed by the SN 1987A bound have a quite rich phenomenology. We presented a comprehensive analysis of this phenomenology in Sections~\ref{sec:4} and \ref{sec:5}, for the case of ALPs coupled respectively with photons or with leptons, besides nucleons. Such ALPs have finite decay lengths, $\lambda_i(m_a,g_{ai})$, controlled by their mass and couplings. Depending on the value of $\lambda_i$, the ALP decay products can deposit energy inside the SN envelope, impacting the explosion mechanism, or they can propagate in the space outside the SN, originating observable signatures. Our results are summarized in Figures~\ref{fig:PhotonBound} and \ref{fig:DSNALPBfluxes}, in the case of a non-vanishing ALP-photon coupling, and in Figures~\ref{fig:gaegap100} and ~\ref{fig:galgap}, for a finite ALP-lepton coupling.

Our study confirms once more the high physics potential of SNe as laboratories for exotic particles. The case of massive ALPs coupled with nucleons has been overlooked so far. We hope that our analysis in this work will stimulate further developments.

\begin{acknowledgments}
We warmly thank Francesca Calore and Georg Raffelt for very useful suggestions during the development of this work. We are grateful to Tobias Fischer for providing us routines used to compute the pion chemical potential and pion abundance.
This article is based upon work from COST Action COSMIC WISPers CA21106, supported by COST (European Cooperation in Science and Technology).
The work of A.L.,  G.L., A.M., 
was partially supported by the research grant number 2017W4HA7S ``NAT-NET: Neutrino and Astroparticle Theory Network'' under the program PRIN 2017 funded by the Italian Ministero dell'Università e della Ricerca (MIUR).
The work of P.C. is supported by the European Research Council under Grant No.~742104 and by the Swedish Research Council (VR) under grants  2018-03641 and 2019-02337. 
\end{acknowledgments}

\bibliographystyle{bibi}
\bibliography{references.bib}

\end{document}